\documentclass[doublecol]{epl2} 

\usepackage{epsf}

\title{Entanglement dynamics and relaxation in a few qubit system interacting with random collisions}

\author{ Giuseppe Gennaro\inst{1} \and  Giuliano Benenti\inst{2}  \and G.Massimo Palma\inst{1}}
\shortauthor{G.Gennaro, G.Benenti and G.M.Palma}

\institute{
 \inst{1} NEST - CNR (INFM) \& Dipartimento  di Scienze Fisiche ed Astronomiche, \\ Universit\`a degli Studi di Palermo, via Archirafi 36,
I-90123 Palermo, Italy     \\         
  \inst{2} CNISM, CNR (INFM) \& Center for Nonlinear and Complex systems,
\\ Universit\`{a} degli Studi dell'Insubria, via Valleggio 11,
I-22100 Como, Italy  \\ \& Istituto Nazionale di Fisica Nucleare, Sezione di Milano, via Celoria 16, I-20133 Milano, Italy
}

\date{\today}

\abstract{The dynamics of a single qubit interacting by a 
sequence of pairwise collisions with an environment consisting 
of just two more qubits is analyzed.  Each collision is modeled in terms of  a random 
unitary operator with a uniform probability distribution described by the uniform Haar measure. We show 
that the purity of the system qubit 
as well as the bipartite and the tripartite entanglement reach time averaged equilibrium 
values characterized by large instantaneous fluctuations.
These equilibrium values are independent of the order of collision among the qubits.
The relaxation to equilibrium is analyzed also in terms of an ensemble average of random collision histories. Such average allows for a quantitative evaluation and interpretation of the decay constants. Furthermore a dependence of the transient dynamics on the initial degree of entanglement between the environment qubits is shown to exist. 
Finally the statistical properties of bipartite and tripartite entanglement are analyzed.
}

\usepackage{amsfonts}

\begin{document}

\maketitle

\section{Introduction}
The repeated collision model has been recently used in literature to analyze the irreversible dynamics of a qubit interacting with a reservoir consisting of a large number of environmental qubits. In particular processes like thermalization \cite{Scarani2002} and homogenization \cite{Ziman2002,ZimanI,ZimanII,ZimanIII}, have been analytically investigated. The same model has ben used recently also to analyze the dynamics of a qubit interacting with a very small environment consisting of just two qubits \cite{Palma2007}. The interest for such system is due to the fact that, at variance with what happens in the case of an environment with a large number of degrees of freedom, the system dynamics cannot be described by a Markovian master equation. Indeed, due to the fact that the system qubit collides repeatedly with the same environment qubits, the dynamics is characterized by large fluctuations and only when the sequence of collision is random a time averaged equilibrium is reached. 
While in all the above mentioned papers the - elastic - collisions have been modeled by a partial swap unitary operator, in the present paper we will analyze the system dynamics in the case in which the two-qubit
collisions are described by random unitary operators 
\cite{Diaconis2005, Mezzadri2007, Kus1991, Kus1994, Kus1996, Kus1998}. 

Random unitary operators have received considerable attention in quantum 
information theory, mainly because they find applications in various 
quantum protocols \cite{Harrow2004,Bennett2004,Hayden2004,Hayden2006}. 
Unfortunately the implementation of 
a random unitary operator acting on the $n$-qubit Hilbert space 
requires a number of elementary quantum gates that is
exponential in the number of qubits. 
On the other hand sequences of random two-qubit gates (collisions)
generate pseudo-random unitary operators
which approximate, to the desired accuracy, the entanglement
properties of true $n$-qubit random states
\cite{Emerson2003, Emerson2004, Emerson2005, Weinstein2005,
Plenio2007,Oliveira2007,Znidaric2007}. 
This approach has given very good results, showing that  pseudo-random
states can be generated efficiently, that is polynomially in $n$
\cite{Emerson2003, Emerson2004, Emerson2005, Weinstein2005,
Plenio2007,Oliveira2007,Znidaric2007}. 

Our choice to describe the pairwise collisions in terms of random two qubit unitary operators is motivated by the 
fact that often a precise modelization 
of the interaction is very hard and that, on the other hand, a good 
description of the approach to equilibrium can be obtained by suitable averages 
of the quantities of interest, as we will describe below.
Furthermore such collision model exhibits interesting
features ranging from memory effects to the efficient entanglement 
generation between system and environment.

\section{The model}
In order to illustrate the approach of our work, we first  
review the repeated collision model. Let us consider a set of $N+1$
qubits, the first of which is the system qubit and the 
remaining $N$ are the reservoir. The system-environment interaction
is due to pairwise  collisions between the system and 
a singe reservoir qubit. 
After $t$ collisions the overall state of the system plus reservoir is
\begin{equation}
\varrho_{SE}^{(t)} = U_{i_t} \cdots U_{i_2} U_{i_1}
\varrho_{SE}^{(0)} U_{i_1}^{\dagger}U_{i_2}^{\dagger} \cdots
U_{i_t}^{\dagger}, \label{Un}
\end{equation}
where $\varrho_{SE}$ is the total density operator and the sequence
$i_1 \cdots i_t$ specify the order with which the
environment qubits collide with the system one. 
As in \cite{Palma2007} we have concentrated our attention to the 
case in which the environment consists of just two qubits.
In the following $0$ will label the system qubit while $1,2$ will label the environment qubits.
At variance with the previous work, however, we have considered here 
the case in which each collision is described by a random 
unitary operator $U_{i}$ picked up from the uniform 
Haar measure on the group $U(4)$. 
In particular in our calculations we have found convenient to parametrize 
each random unitary matrix $U_i$, in terms of  the Hurwitz representation 
of the unitary group $U(4)$ \cite{Hurwitz1897,Kus1998,Weinstein2005}.
Such different choice of the collision unitary operators has several 
consequences. First of all the collisions are of course no longer elastic 
and, regardless of the number of environment qubits, homogenization 
is no longer achieved. However, even  for a few qubits environment,  
a time averaged equilibrium state with large fluctuations is reached 
regardless of the order with which the qubits collide (in the numerical
data shown below, we consider random sequences $i_1,\cdots i_t$). 
Furthermore such equilibrium state is independent of the initial state 
of the qubits.
In the following we will characterize some aspects of such approach 
to a time averaged equilibrium state.

Random qubit-environment interactions have been
recently considered \cite{Pineda2007,Petruccione2007}, for
high dimensional environments. However  in our work we deal with
small environments, such that the information acquired by the environment on 
the system can flow back to the system and a Markovian description of
our model is surely not possible. Moreover, there is no weak coupling
parameter and the state of the environment is significantly 
affected by the interaction at each collision,
so that also the Born approximation does not apply.

\section{Approach to equilibrium} 

\begin{figure}[htbp!]
\begin{center}
\includegraphics[height=3.2cm]{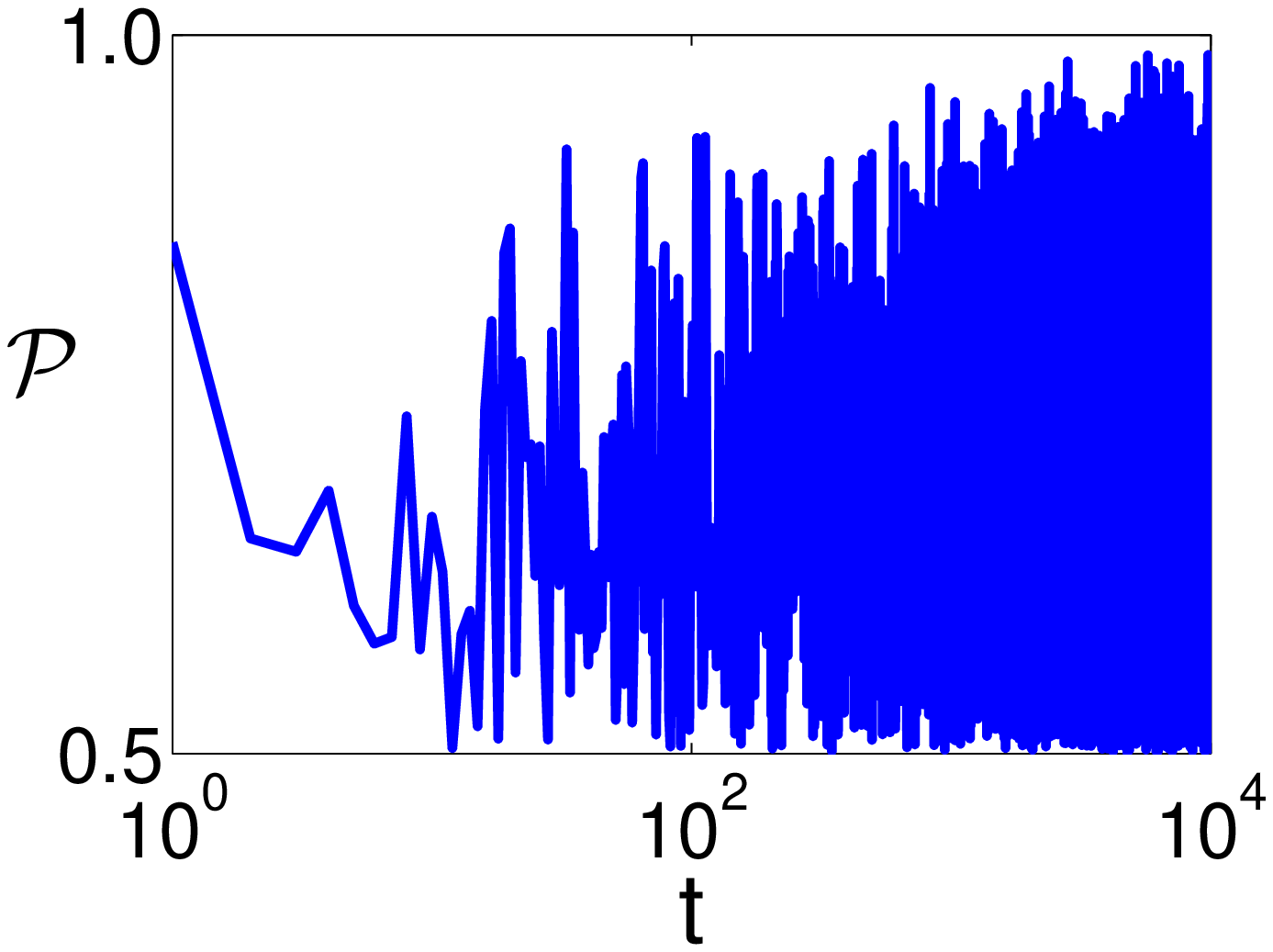}
\includegraphics[height=3.2cm]{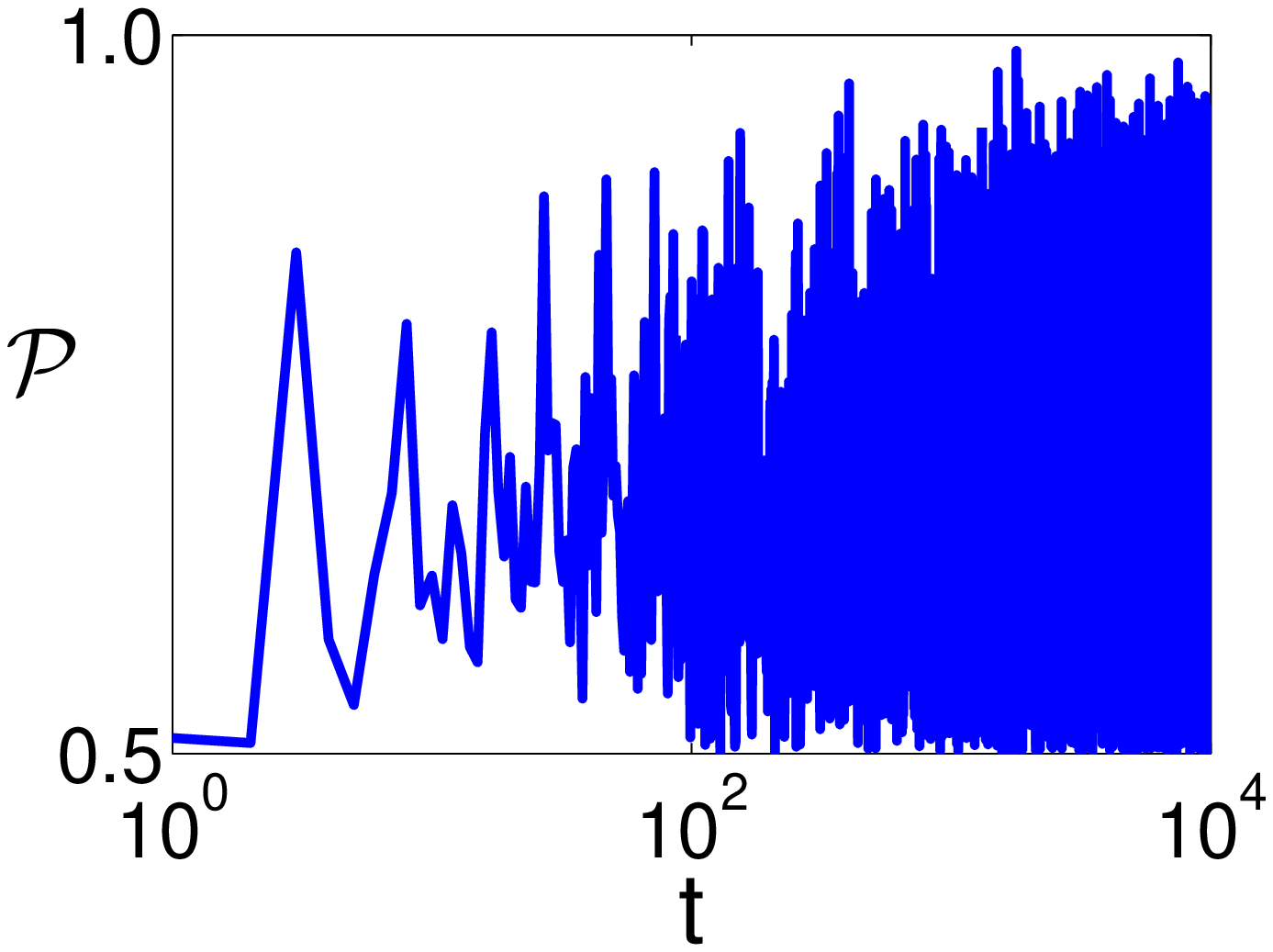}
\\
\includegraphics[height=3.2cm]{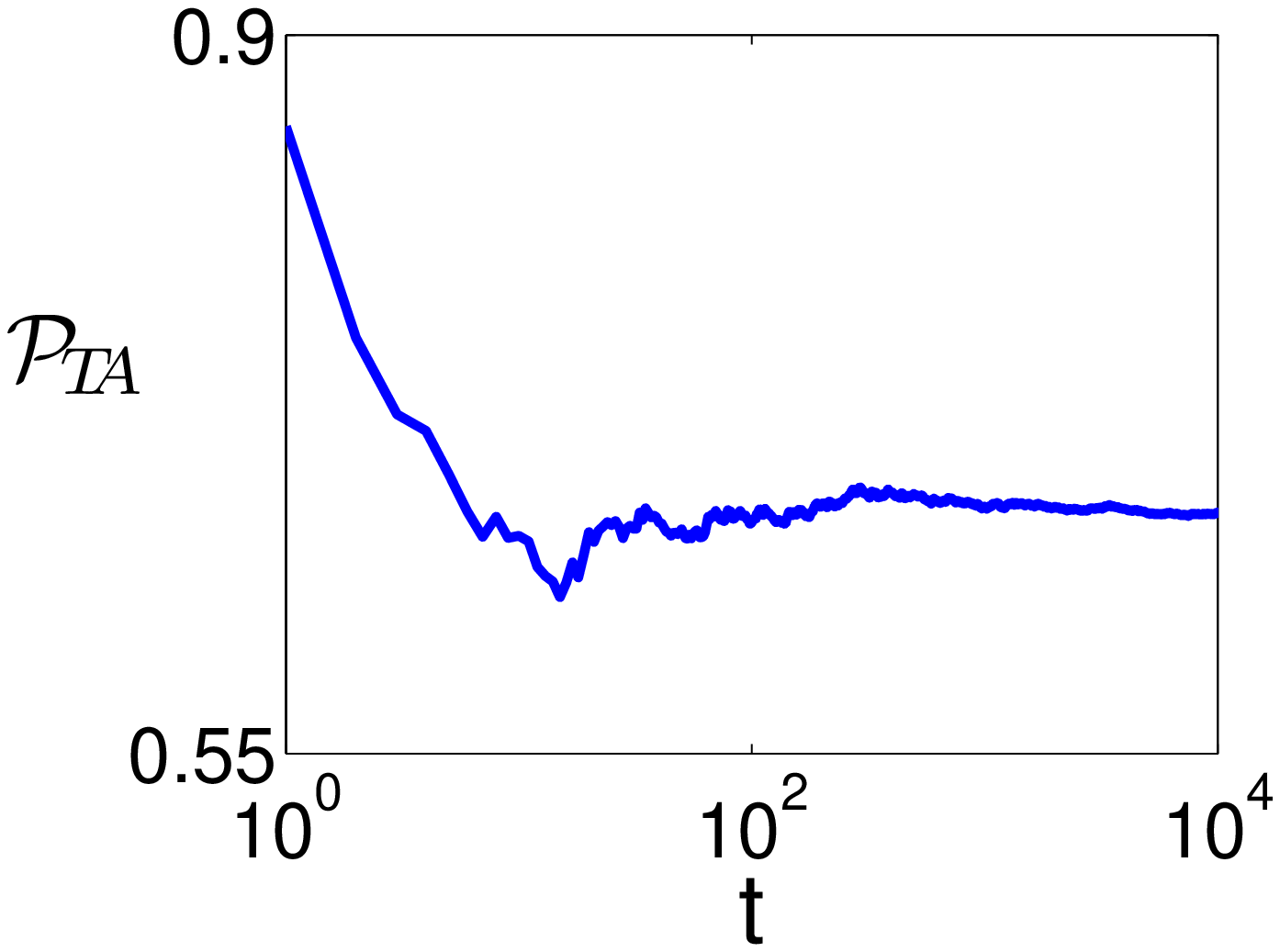}
\includegraphics[height=3.2cm]{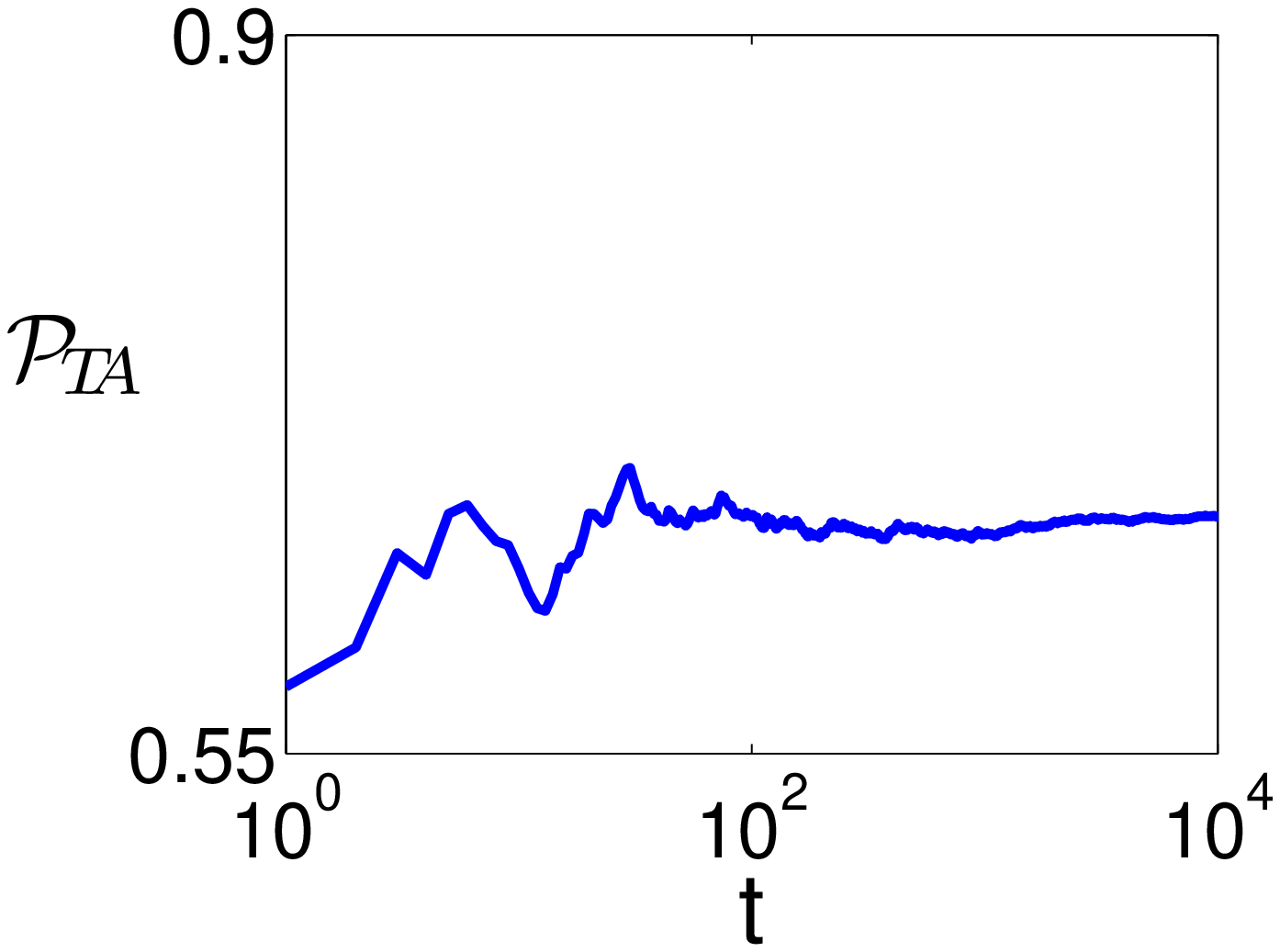}
\\
\includegraphics[height=3.2cm]{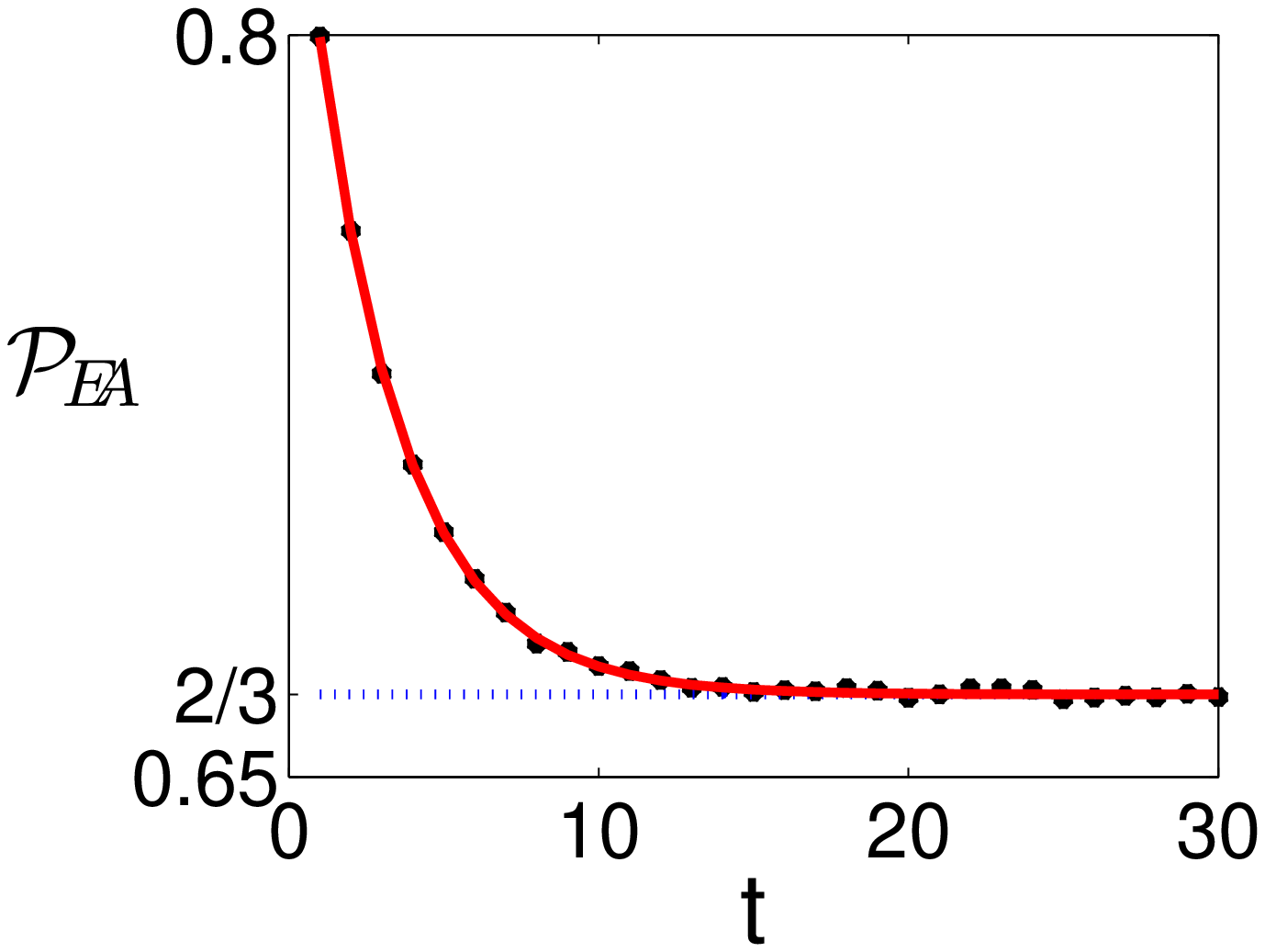}
\includegraphics[height=3.2cm]{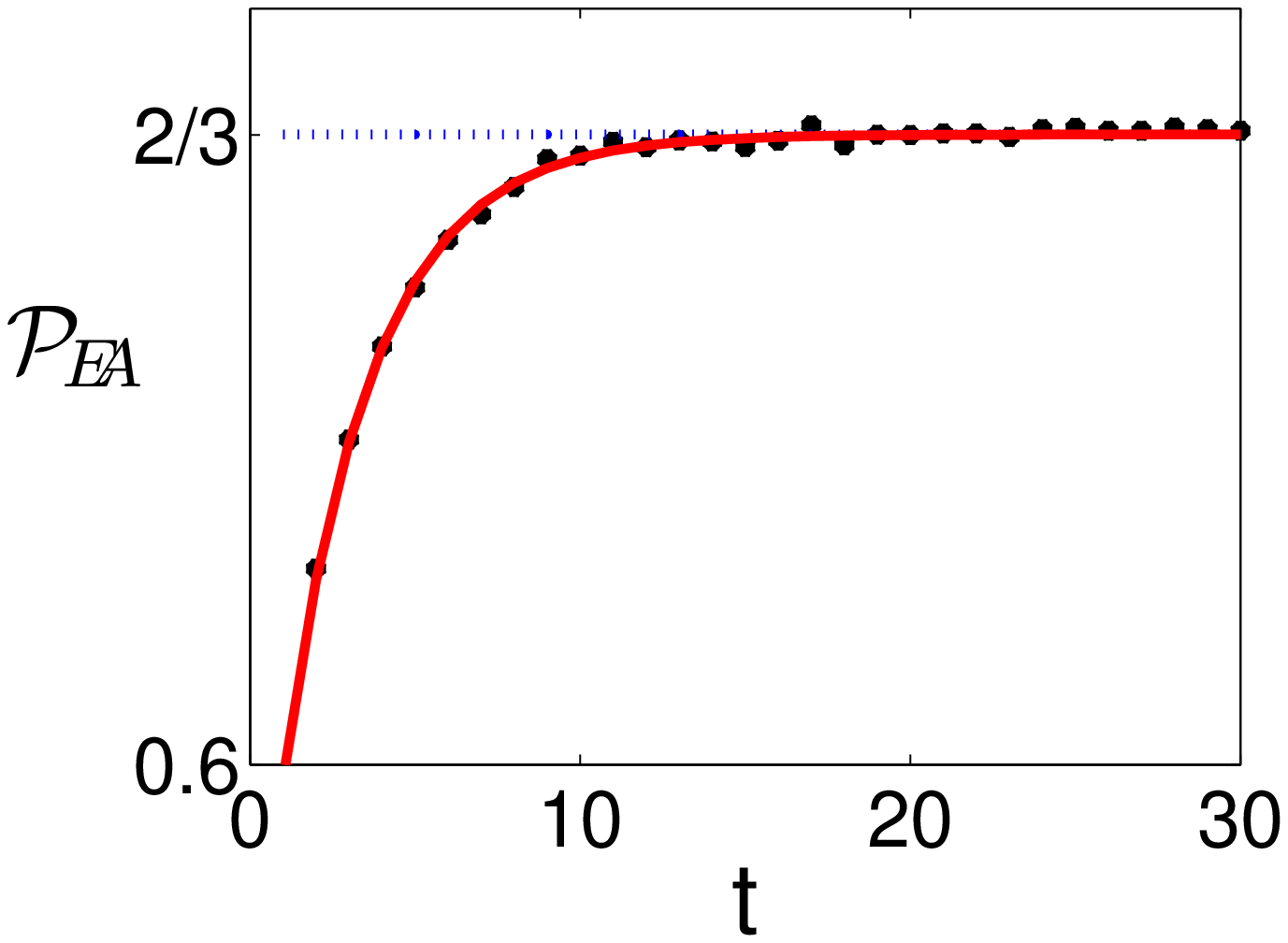}
\end{center}
\caption{From top to bottom: instantaneous, time averaged and 
ensemble averaged purity, for an environment in an 
initial product (left) or maximally entangled (right) state.
The time evolution of the ensemble averaged purity is
in both cases well fitted by the exponential curve 
$\mathcal{P}_{EA}(t) - \mathcal{P}_{L}\propto \exp(-0.36 t)$,
with $\mathcal{P}_L=\frac{2}{3}$ (asymptotic horizontal line).} 
\label{fig:purity}
\end{figure}

We have characterized the decoherence of the system qubit  
in terms of the purity $\mathcal{P}$. We remind the reader
that the purity is defined as  
$\mathcal{P}=\hbox{Tr}\left[{\varrho_{S}^{2}}\right]$,
where $\varrho_{S} = \hbox{Tr}_E\left[\varrho_{S E}\right]$
is the reduced density operator of the system qubit. The purity is a decreasing function of the degree of statistical mixture
of the qubit and takes values in the range
$ \frac{1}{2} \leq\mathcal{P}\leq 1$
where $ \mathcal{P}=1$ corresponds to pure states
and $\mathcal{P}=\frac{1}{2}$ to the completely unpolarized mixed state. 
As already mentioned, due to the small number of  environment qubits, 
the  instantaneous system purity undergoes large
fluctuations regardless of the initial state of
both the system and the environment and regardless of the
sequence of collisions as shown in Fig.~\ref{fig:purity} (first row). 
The purity, however, approaches a time 
averaged equilibrium state. 
To see this we calculate the time averaged purity 
$\mathcal{P}_{TA}(t)  $ as
\begin{equation}
\mathcal{P}_{TA}(t) = \frac{1}{t+1} \sum_{t^\prime=0}^t 
\hbox{Tr}\left[\rho_{S}^{2}(t')\right].
\label{TA}
\end{equation}
As shown in Fig.~\ref{fig:purity} (second row),  
$\mathcal{P}_{TA}(t)$ reaches the same equilibrium value  
regardless of the initial entanglement of the environment qubits, 
a natural consequence of the random nature of the collisions.

Further insight into the approach to equilibrium of the system qubit is 
obtained if rather than the time average purity we evaluate the  
ensemble averaged purity  $\mathcal{P}_{EA}(t)$ over the uniform 
Haar measure. As we shall see, an analysis of the time evolution of 
such averaged 
quantity allows us to determine the time scale for
the relaxation to equilibrium. 
To evaluate $\mathcal{P}_{EA}(t)$  we have generated 
a large number of sequences of random collisions, each one 
drawn from the uniform measure, and we have averaged  the purity
over the different histories after $t$ steps.
Also such ensemble average shows  an irreversible behaviour of 
the qubit dynamics. Regardless of the initial state of the environment 
qubits, $\mathcal{P}_{EA}(t)$ decays to the value $\frac{2}{3}$, which 
coincides with the average purity 
$\mathcal{P}_{L}$ predicted by Lubkin  \cite{Lubkin1978}
for true overall (system-environment) random states. 
Note that $\mathcal{P}_{TA}(t)$ and $\mathcal{P}_{EA}(t)$ tend to the same asymptotic value.
If $\mu$ and $\nu$ are respectively the dimensions of the Hilbert 
space of the system and of the environment one has
 \begin{equation}
\mathcal{P}_L=\frac{\mu+\nu}{\mu\nu+1}.
\end{equation}
In our case $\mu=2$ and $\nu=4$, and therefore $\mathcal{P}_L =\frac{2}{3}$.  
However, as shown in Fig.~\ref{fig:purity} (third row) the time
evolution of
$\mathcal{P}_{EA}(t)$ exhibits a clear dependence on the degree of entanglement of the initial 
state  of the environment qubits. 
Notably the exponential approach to 
equilibrium is for the first collisions 
a decreasing (increasing) function of $t$ for an environment 
in an initial product (maximally entangled) state.
In both cases a numerical fit shows that 
$\mathcal{P}_{EA}(t)- \mathcal{P}_L \propto 
\exp(-\lambda t)$ with $\lambda\approx 0.36$.

The asymptotic relaxation to equilibrium can be computed analytically 
following the approach developed in 
\cite{Plenio2007,Oliveira2007,Znidaric2007}. Each pure three-qubit
state $\rho_{SE}$ can be expanded over products of Pauli matrices:
$\rho_{SE}=\sum_{\alpha_0,\alpha_1,\alpha_2}c_{\alpha_0\alpha_1\alpha_2}
\sigma_0^{\alpha_0}\otimes \sigma_1^{\alpha_1}\otimes \sigma_2^{\alpha_2}$,
where $\sigma_i^{\alpha_i}$ denotes a Pauli matrix
acting on the $i$th qubit, with  $\alpha_i\in\{0,x,y,z\}$ and  $\sigma^0=I$. 
The purity then reads
$\mathcal{P}(t)=\sum_{\alpha_0} c_{\alpha_000}^2(t)$. The purity decay 
can therefore be obtained from the evolution in time of the coefficients
$c_{\alpha_000}^2$. For the random collision model the column vector
$c^2$ of the coefficients $c_{\alpha_0\alpha_1\alpha_2}^2$ evolves 
according to a Markov chain dynamics of the form $c^2(t+1)=Mc^2(t)$ 
\cite{Znidaric2007}. The matrix $M$ is obtained after averaging over
the two possible couplings $01$ and $02$: $M=\frac{1}{2}(M^{(2)}_{01}+
M^{(2)}_{02})$, with $M^{(2)}_{ij}$ acting non trivially 
(differently from identity) only on the subspace spanned
by qubits $i$ and $j$. Averaging over the uniform Haar measure on 
$U(4)$ one can see that $M^{(2)}_{ij}$ preserves identity 
($\sigma_i^0\otimes \sigma_j^0\to \sigma_i^0\otimes \sigma_j^0$)
and uniformly mixes the other $15$ products $\sigma_i^{\alpha_i}\otimes
\sigma_j^{\alpha_j}$ \cite{Znidaric2007}. The matrix $M$ has an
eigenvalue equal to $1$ (with multiplicity $2$) and all the other eigenvalues
smaller than $1$. Therefore the asymptotic purity decay is determined
by the gap $\Delta$ in the Markov chain, namely by 
the second largest eigenvalue $1-\Delta$ of the matrix $M$.
We have $\mathcal{P}_{EA}(t)-\mathcal{P}_L\asymp (1-\Delta)^t=
\exp\{[\ln(1-\Delta)]t\}$. In our model $1-\Delta=0.7$ 
(eigenvalue with multiplicity $2$) and therefore 
$-\ln(1-\Delta)\approx 0.357$, in very good agreement with the 
value $\lambda\approx 0.36$ obtained from our fit. 

\section{Entanglement dynamics} 

The dynamics of bipartite and multipartite entanglement between the qubits of our model
shows interesting features.
Such dynamics has been conveniently characterized in terms of the concurrence and 
of the tangles \cite{Wootters1998,Wootters2000}. We remind the reader that, 
given the density operator $\rho_{ij}$ of a
bipartite system of two qubits, the tangle $\tau_{i|j}$ is defined as
\begin{equation}
\tau_{i|j}(\rho)=[\max\left\{0,\alpha_{1}-\alpha_{2}-{\alpha_{3}}-{\alpha_{4}}\right\}]^2,
\end{equation}
where $\left\{\alpha_{k}\right\}$ ($k=1,..,4$) are the square roots of the eigenvalues (in non-increasing order) of the
non-Hermitian operator $\bar{\rho}_{ij}=
\rho_{ij}(\sigma_{y}\otimes\sigma_{y})
\rho_{ij}^{*}(\sigma_{y}\otimes\sigma_{y})$, $\sigma_{y}$ is the
$y$-Pauli operator and $\rho_{ij}^{*}$ is the complex conjugate 
of $\rho_{ij}$, in the eigenbasis of the $\sigma_z  \otimes\sigma_z$ operator. 
The concurrence $C$ is defined simply as $C_{ij}=\sqrt{\tau_{i|j}}$.
The tangle $\tau_{i|j}$, or equivalently the concurrence 
$C_{ij}$ can be used to quantify the entanglement between the pair of 
qubits $i,j$ for an arbitrary
reduced density operator $\rho_{ij}$.  Furthermore, when the overall 
state of the system is pure, the amount of entanglement
between qubit $i$ and all the remaining can be quantified by the 
tangle $\tau_{i| \mbox{rest}}=4\det\rho_i$. 
The tangle $\tau_{0|\mbox{rest}}$ between the system qubit and the environment
conveys the same information as the purity $\mathcal P$. Indeed it is easy to show that
 $\tau_{0|\mbox{rest}}=2-2{\mathcal P}$.
We have numerically
computed the tangles $\tau_{0|1}$, $\tau_{0|2}$, and 
$\tau_{1|2}$ of the two-qubit reduced density matrices and the three-tangle
$\tau_{i|j|k}=\tau_{i|jk}-\tau_{i|j} - \tau_{i|k}$, where 
$i,j,k$ can be any permutation of $0,1,2$ and where the tangle $\tau_{i|jk}$
measures the entanglement between the $i$th qubit and the rest of the system, 
i.e., qubits $j,k$. The three-tangle $\tau_{0|1|2}$ is
a measure of the purely tripartite entanglement and is invariant under 
permutations of the three qubits \cite{Wootters2000}.

The instantaneous dynamics  of the tangles $\tau_{i|j}$
and of the three-tangle $\tau_{0|1|2}$ is similar to the dynamics 
shown in the Fig.~\ref{fig:purity} (top) for the purity, that is,
these quantities are characterized by large instantaneous fluctuations. 
However the time averaged tangles $(\tau_{i|j})_{TA}(t)$, defined in analogy with Eq.(\ref{TA}), 
approach the same limiting value $\tau_P$.  

\begin{figure}[htpb!]
\includegraphics[height=3.2cm]{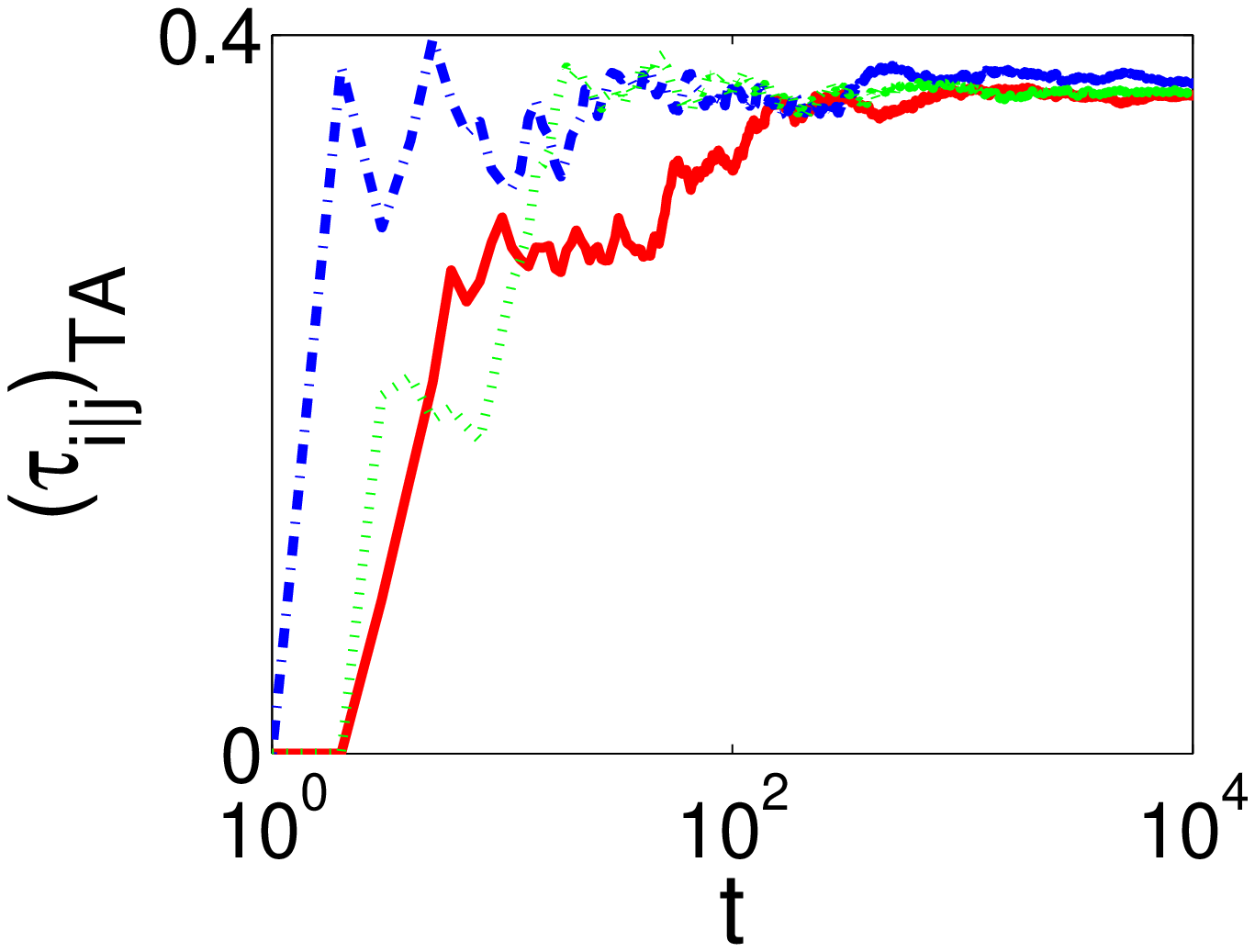}
\includegraphics[height=3.2cm]{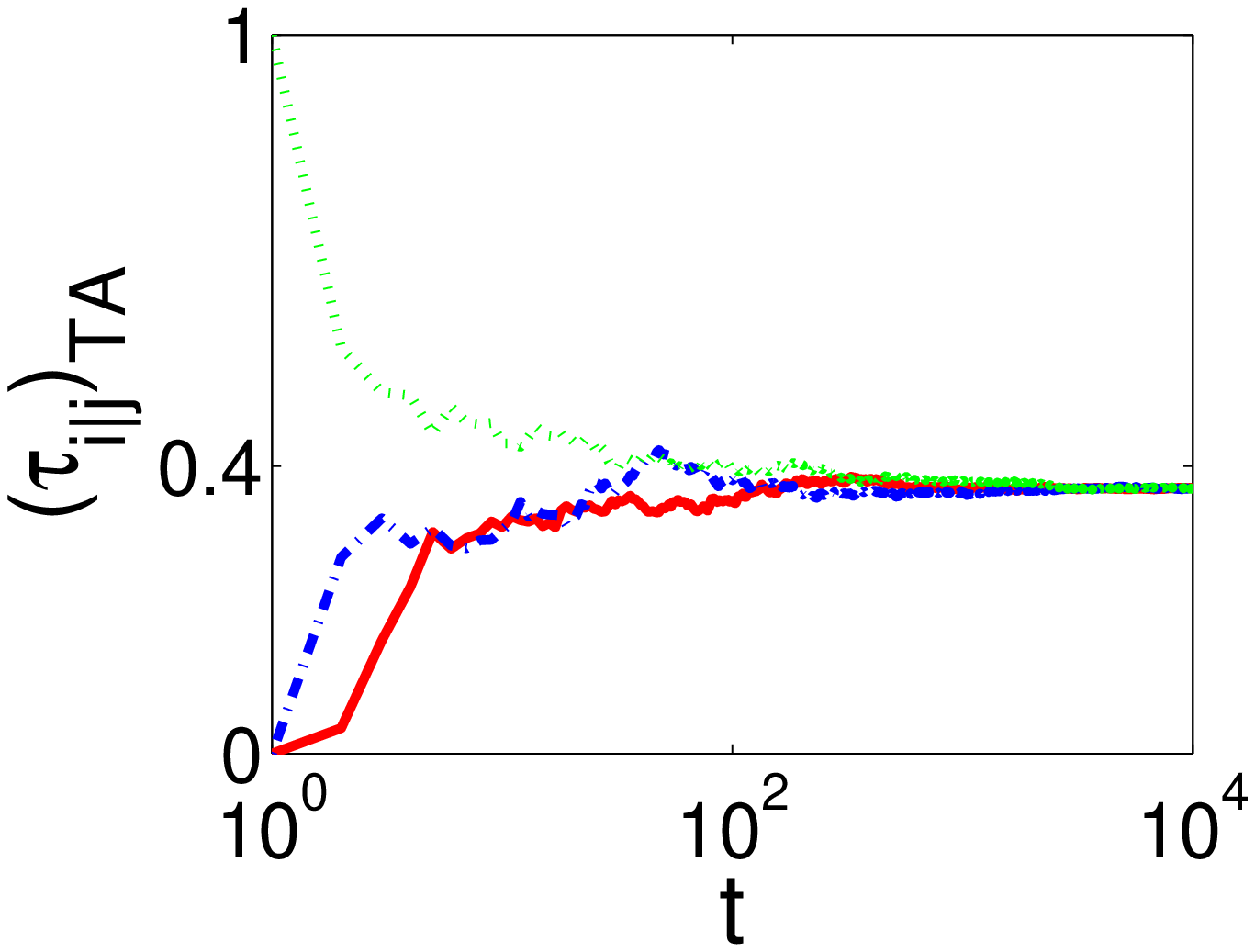}
\\
\includegraphics[height=3.2cm]{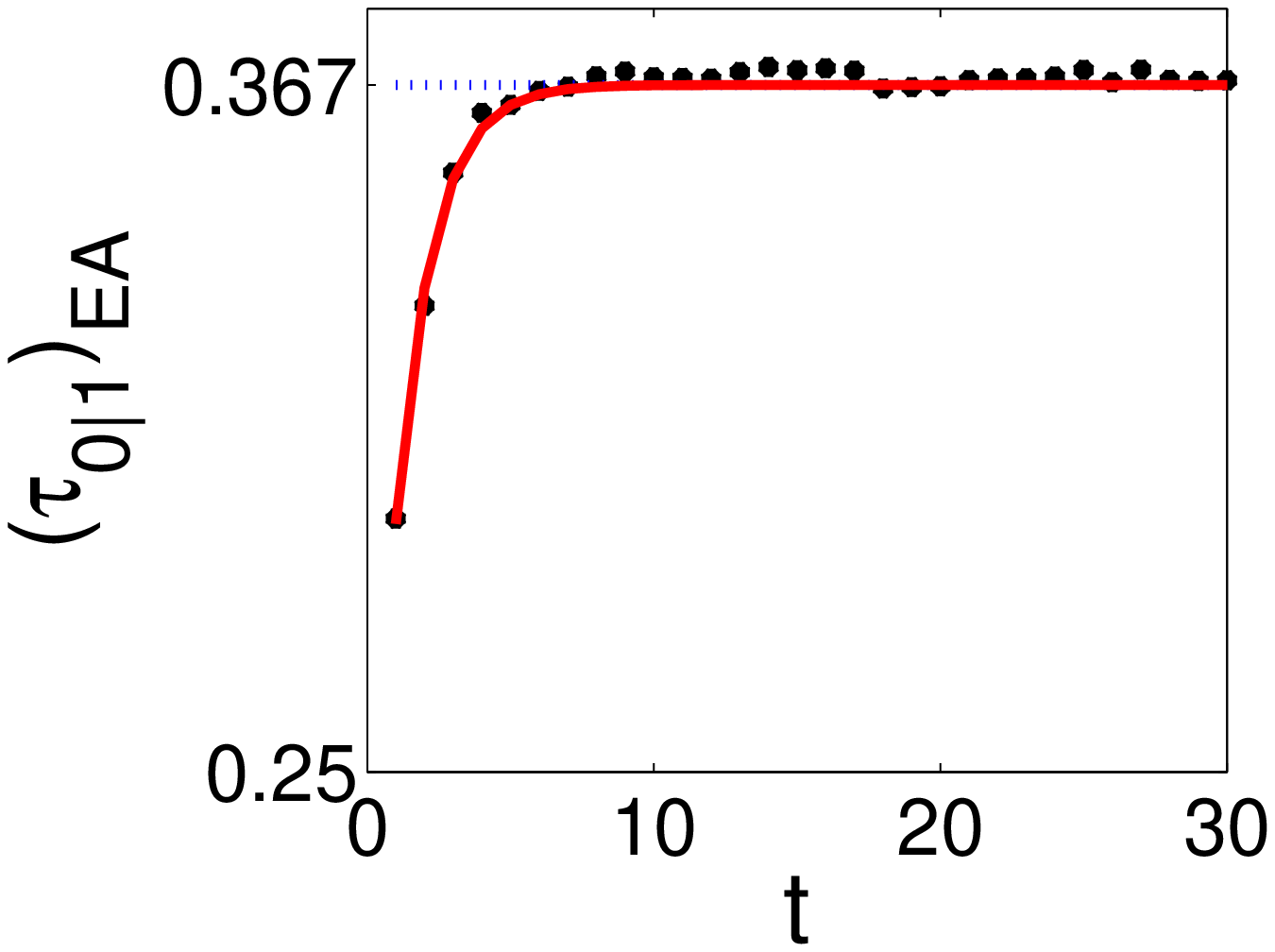}
\includegraphics[height=3.2cm]{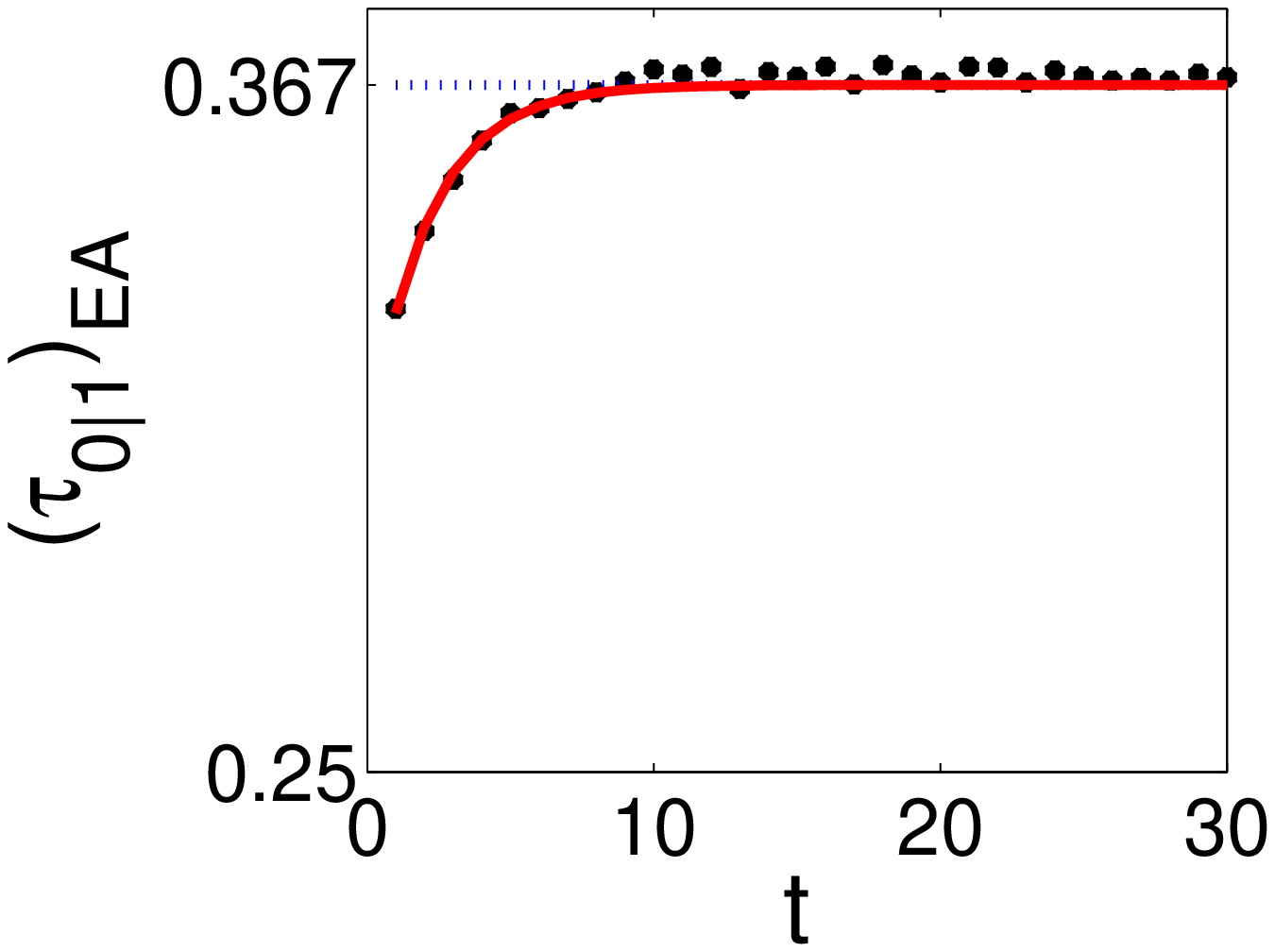}
\\
\includegraphics[height=3.2cm]{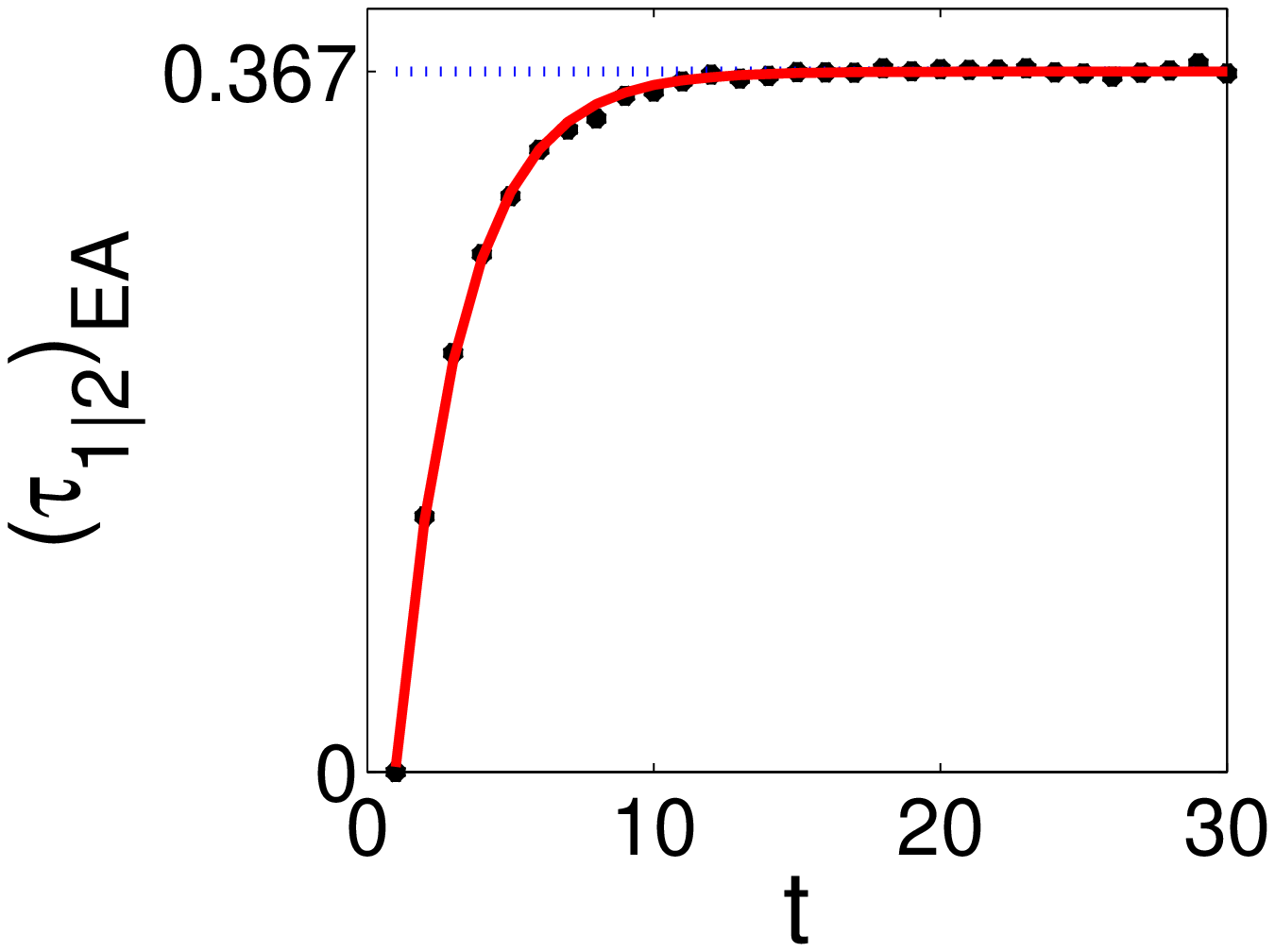}
\includegraphics[height=3.2cm]{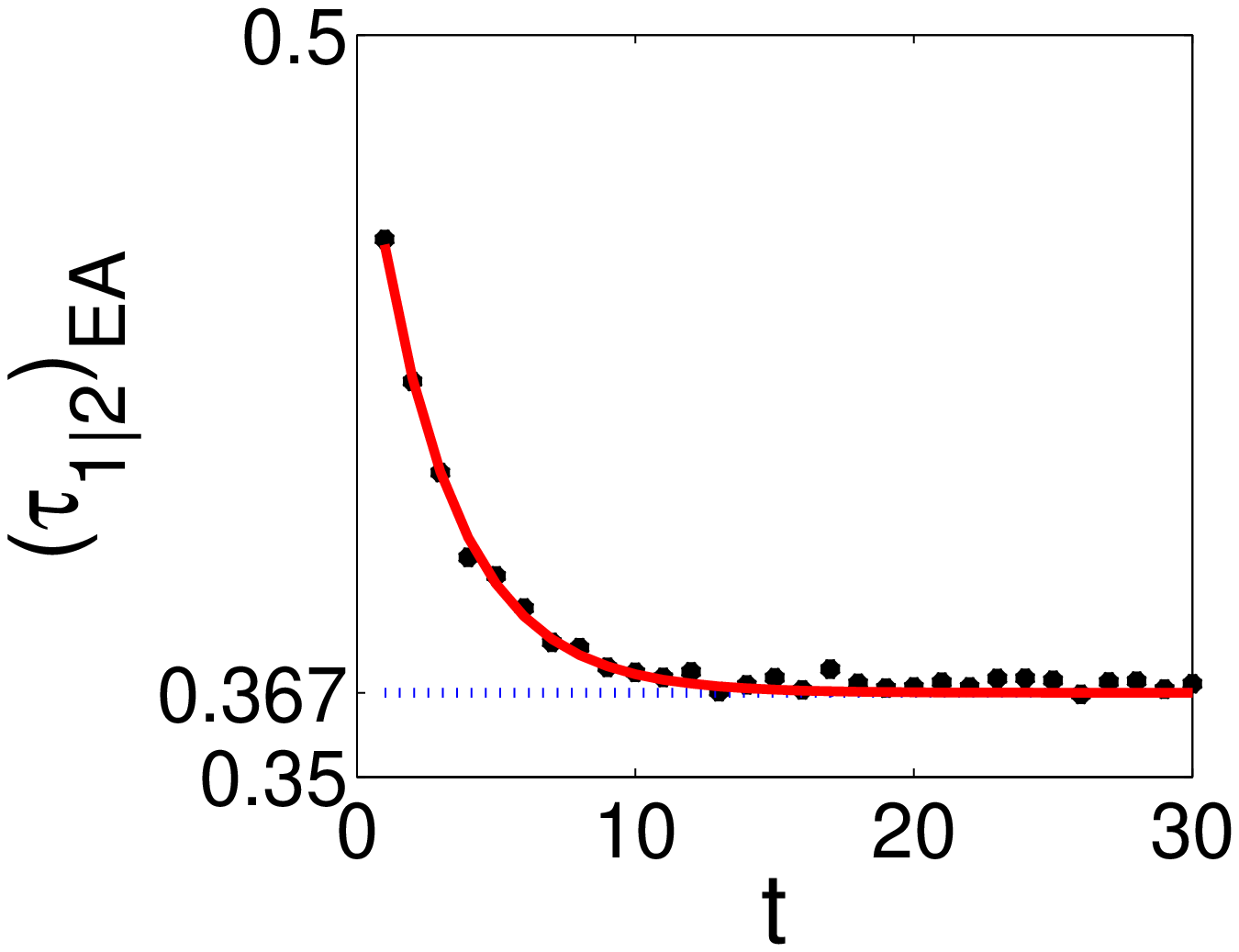}
\caption{Time (first row) and ensemble (second and third rows) averages for the 
pairwise tangles $\tau_{0|1}$ (solid line in the first row), $\tau_{0|2}$ (dashed line in the first row) 
and $\tau_{1|2}$ (dotted line in the first row), for an environment in an initial
product (left) or maximally entangled (right) state.
For the ensemble averaged quantities also the fit
$(\tau_{i|j})_{EA}(t) - \tau_{P}\propto \exp(-\lambda_{ij} t)$
is shown. In the second row only the $(\tau_{0|1})_{EA}(t)$ is shown due to the same figure of $(\tau_{0|2})_{EA}(t)$  }
\label{fig:concurrence}
\end{figure}

Again we have numerically evaluated the ensemble 
average (with respect to the Haar measure) tangles $(\tau_{i|j})_{EA}(t)$. As shown in Fig.~\ref{fig:concurrence} (bottom) 
the pairwise tangles $(\tau_{0|1})_{EA}(t)$ and  $(\tau_{0|2})_{EA}(t)$ approach exponentially  the same equilibrium value $\tau_P\approx 0.367$. 
The numerical data in 
Fig.~\ref{fig:concurrence} (bottom) are well fitted by the 
curves $(\tau_{i|j})_{EA}(t)-\tau_P\propto \exp(-\lambda_{ij}t)$.
When the environment is initially separable we obtain 
$\lambda_{01}\approx \lambda_{02}\approx 0.76$, $\lambda_{12}\approx 0.44$,
while for an initially maximally entangled states of the environment we have
$\lambda_{01}\approx \lambda_{02}\approx 0.44$, $\lambda_{12}\approx 0.36$.
Therefore, initially entangled environment qubits limit the rate of 
generation of bipartite entanglement between the qubit system and 
a single environment qubit, even though the amount of pairwise entanglement 
obtained asymptotically is always the same.
We have furthermore verified also that the average tangle $(\tau_{1|2})_{EA}$ approaches the same limiting value $\tau_P$ even though the environment  qubits do not collide directly.

\begin{figure}[htbp!]
\begin{center}
\includegraphics[height=3.2cm]{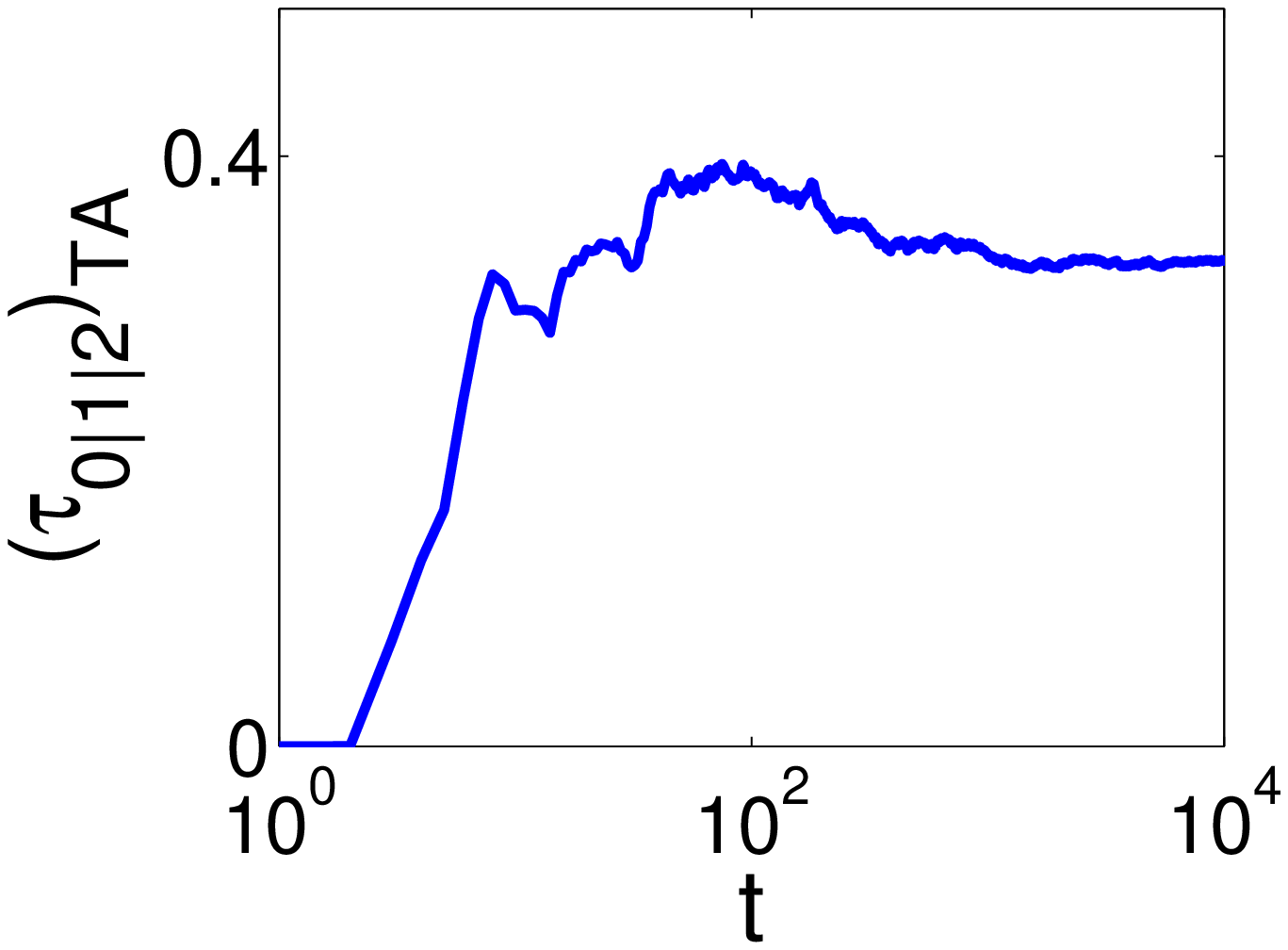}
\includegraphics[height=3.2cm]{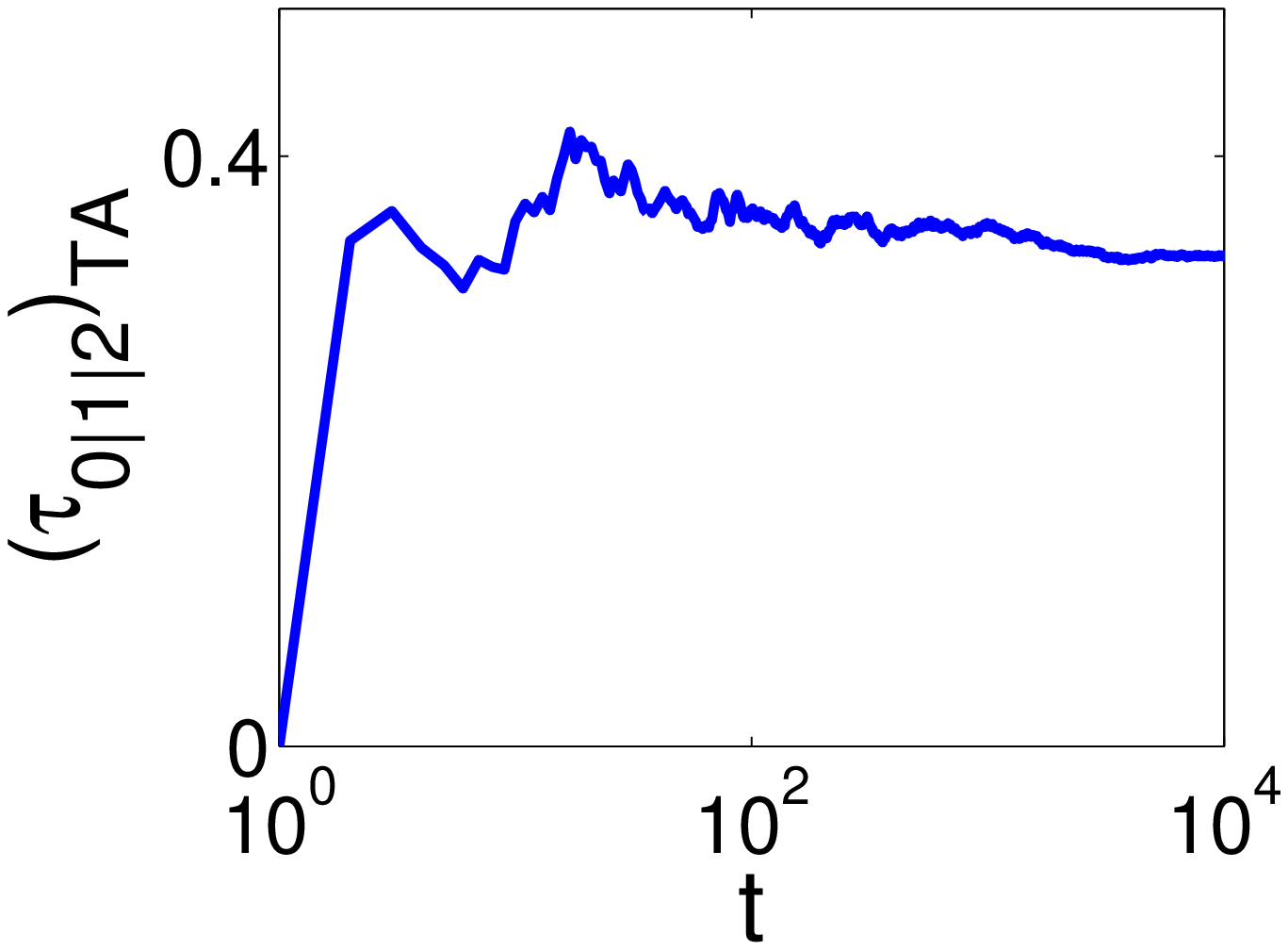}
\\
\includegraphics[height=3.2cm]{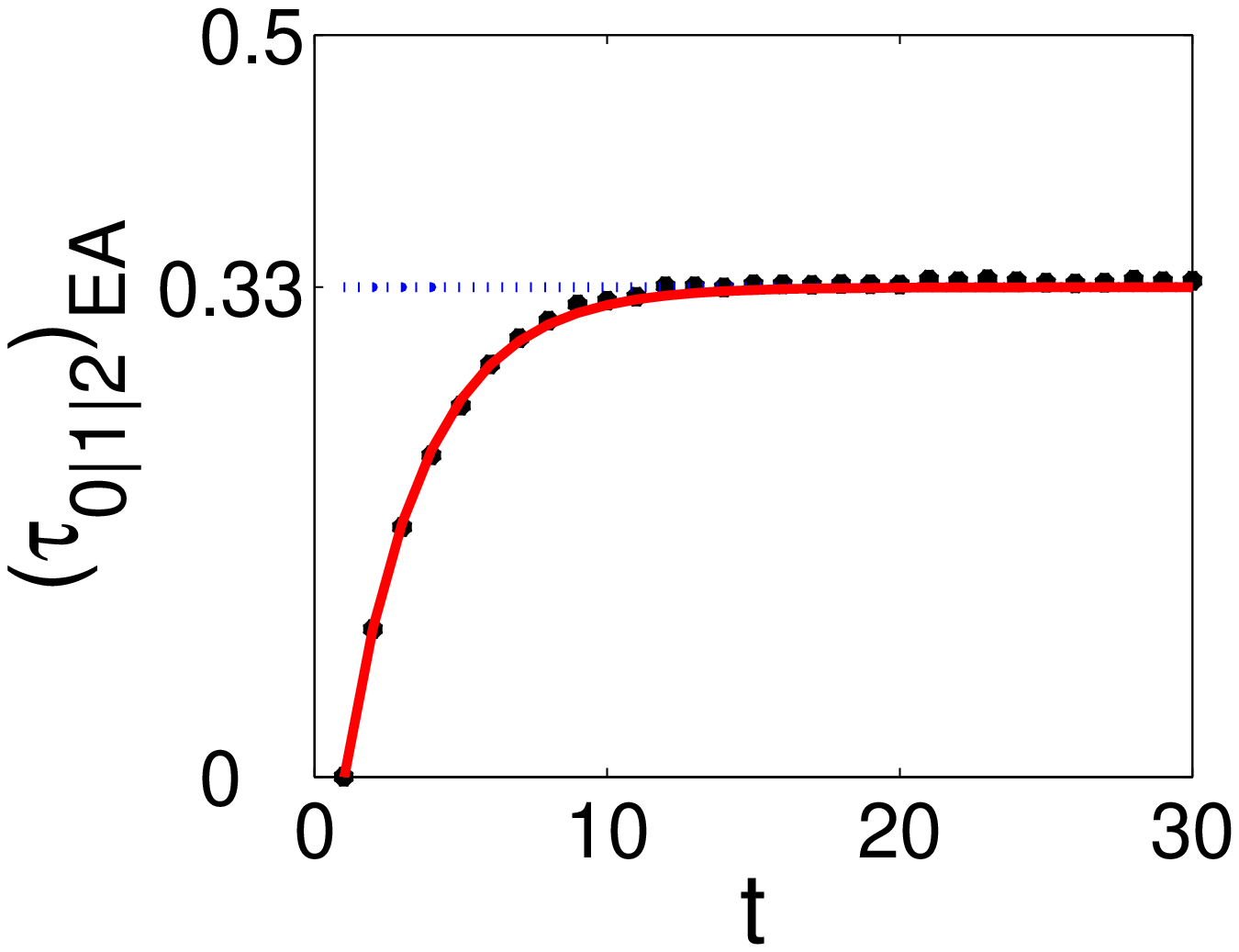}
\includegraphics[height=3.2cm]{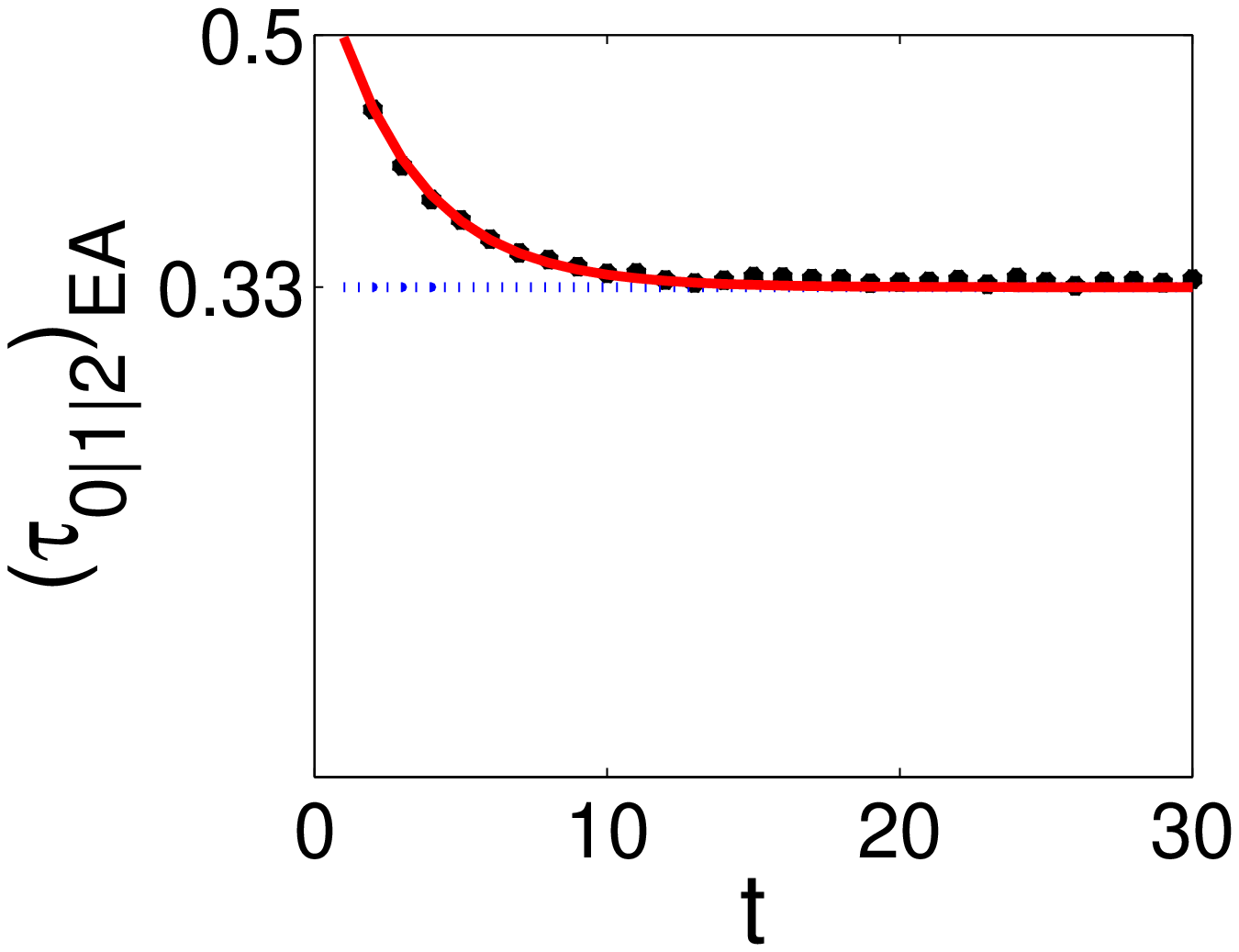}
\end{center}
\caption{Time (top) and ensemble (bottom) averages for the 
three-tangle $\tau_{0|1|2}$ for an environment in an initial
product (left) or maximally entangled (right) state.
For the ensemble averaged quantities also the fit
$(\tau_{0|1|2})_{EA}(t) - \tau_{T}
\propto \exp(-0.36t)$ is shown.}
\label{fig:tangle}
\end{figure}

As one would expect from the above discussion, also the mutipartite entanglement approaches an equilibrium value.  This can be seen by the time and ensemble averages three-tangle $\tau_{0|1|2}$ shown in Fig.~\ref{fig:tangle}.
The approach to the 
equilibrium value $\tau_T$ is exponential also for this quantity and 
we can extract the convergence rate from the fit  
$(\tau_{0|1|2})_{EA}(t)-\tau_T\propto \exp(-\lambda_{012}t)$,
with $\lambda_{012}=0.36$ regardless on whether the initial state of the environment is entangled or separable.

\begin{figure}[htbp!]
\begin{center}
\includegraphics[height=3.2cm]{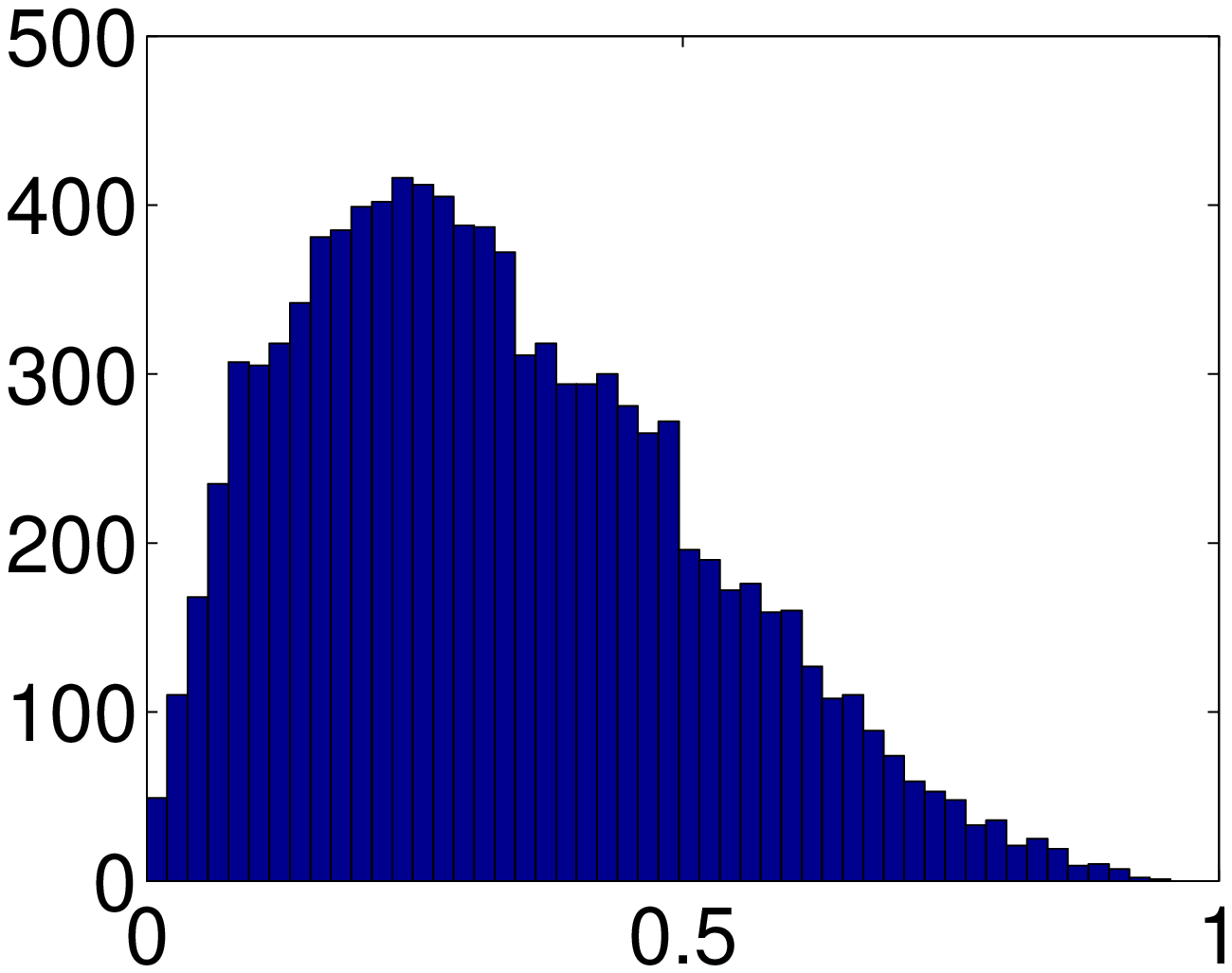}
\includegraphics[height=3.2cm]{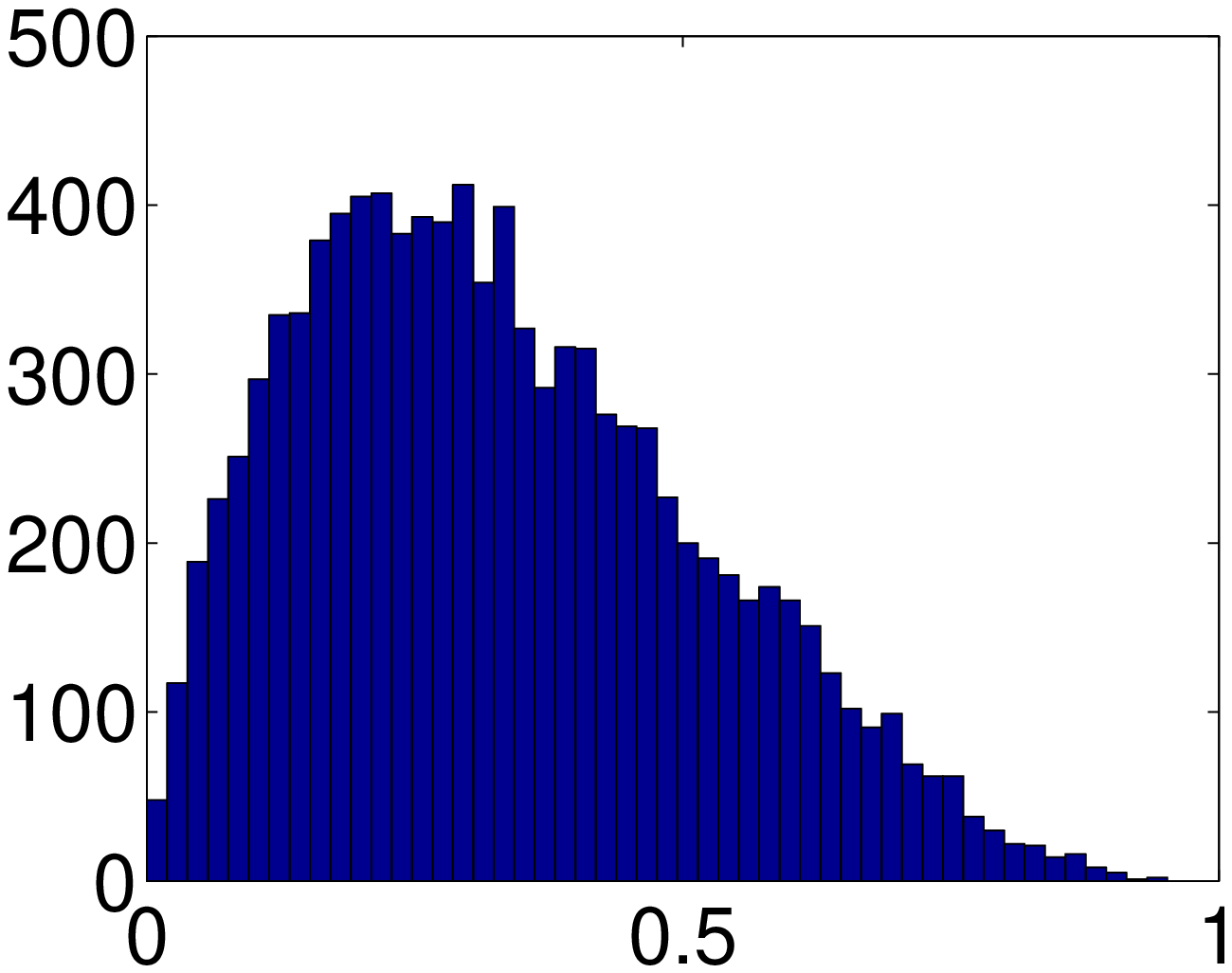}
\end{center}
\caption{Statistical distributions of the three-tangle $\tau_{0|1|2}$, 
for separable (left) and entangled (right) initial state of the environment.
In both cases $\tau_{0|1|2}=0.33\pm 0.19$.}
\label{fig:histtangle}
\end{figure}

Finally, we note that not only the time and ensemble averages 
of the various tangles converge to the same limit values regardless of the 
initial conditions, but also to a well defined limit distribution.
In particular the concurrences are distributed in accordance with the distribution shown in 
in Fig.3 of \cite{Scott2003} for random pure $3$-qubit states.
Finally the numerically calculated distributions of the three-tangle 
in our collision model are shown in Fig.~\ref{fig:histtangle}, for 
separable and for entangled initial environment state.  

\section{Conclusions}
In summary, we have shown that relaxation (in time average) to 
statistical equilibrium is possible for a system of just three qubits 
undergoing purely random unitary evolution, regardless of the order 
of the collisions (in contrast to the case studied in \cite{Palma2007}). 
This process is intrinsically irreversible 
(in contrast to \cite{Scarani2002,Ziman2002}) 
due to the random nature of the interactions. The purity of the qubit 
system shows an exponential decay and the decay rate 
has been evaluated both numerically and analytically.
The limit value is in accordance with \cite{Lubkin1978}.
The limit values (in a time average sense) of  the pairwise 
qubit tangles are the same for all possible qubit pairs.
In contrast to \cite{Palma2007}, residual entanglement (three-tangle) 
is generated regardless of initial entanglement of the environment 
qubits. 

\acknowledgments
G.B. acknowledges support from the PRIN 2005 "Quantum computation with trapped particle arrays, neutral and
charged". G.M.P. and G.G. acknowledge support from the PRIN 2006 "Quantum noise in mesoscopic systems".  
G.B. acknowledges useful discussions with Marko \v Znidari\v c.

\end{document}